\documentclass[aps,prl,reprint,superscriptaddress]{revtex4-1}

\usepackage{graphicx}
\usepackage{amsmath}
\usepackage{latexsym}
\usepackage{textcomp}
\usepackage[colorlinks,linkcolor=blue,citecolor=blue,urlcolor=blue]{hyperref}

\def\degree{${}^{\circ}$}

\begin{document}


\title{Grain Boundaries in Chemical Vapor Deposited Atomically Thin Hexagonal Boron Nitride}


\author{Xibiao Ren}
\affiliation{State Key Laboratory of Silicon Materials, School of Materials Science and Engineering, Zhejiang University, Hangzhou, Zhejiang 310027 , P. R. China}
\author{Jichen Dong} 
\affiliation{Centre for Multidimensional Carbon Materials, Institute for Basic Science, Ulsan 44919, Republic of Korea}
\author{Peng Yang}
\affiliation{State Key Laboratory of Functional Materials for Informatics, Shanghai Institute of Microsystem and Information Technology, Chinese Academy of Sciences, 865 Changning Road, Shanghai 200050, P.R. China}
\author{Jidong Li}
\affiliation{State Key Laboratory of Mechanics and Control of Mechanical Structures, the Key Laboratory of Intelligent Nano Materials and Devices of DoE, Institute of Nano Science of Nanjing University of Aeronautics and Astronautics, Nanjing 210016, China}
\author{Guangyuan Lu}
\affiliation{State Key Laboratory of Functional Materials for Informatics, Shanghai Institute of Microsystem and Information Technology, Chinese Academy of Sciences, 865 Changning Road, Shanghai 200050, P.R. China}
\author{Tianru Wu}
\affiliation{State Key Laboratory of Functional Materials for Informatics, Shanghai Institute of Microsystem and Information Technology, Chinese Academy of Sciences, 865 Changning Road, Shanghai 200050, P.R. China}
\author{Haomin Wang}
\affiliation{State Key Laboratory of Functional Materials for Informatics, Shanghai Institute of Microsystem and Information Technology, Chinese Academy of Sciences, 865 Changning Road, Shanghai 200050, P.R. China}
\author{Wanlin Guo}
\affiliation{State Key Laboratory of Mechanics and Control of Mechanical Structures, the Key Laboratory of Intelligent Nano Materials and Devices of DoE, Institute of Nano Science of Nanjing University of Aeronautics and Astronautics, Nanjing 210016, China}
\author{Ze Zhang}
\affiliation{State Key Laboratory of Silicon Materials, School of Materials Science and Engineering, Zhejiang University, Hangzhou, Zhejiang 310027 , P. R. China}
\author{Feng Ding}
\email[]{f.ding@unist.ac.kr}
\affiliation{Centre for Multidimensional Carbon Materials, Institute for Basic Science, Ulsan 44919, Republic of Korea}
\affiliation{School of Materials Science and Engineering, Ulsan National Institute of Science and Technology (UNIST), Ulsan 44919, Republic of Korea}
\author{Chuanhong Jin}
\email[]{chhjin@zju.edu.cn}
\affiliation{State Key Laboratory of Silicon Materials, School of Materials Science and Engineering, Zhejiang University, Hangzhou, Zhejiang 310027 , P. R. China}


\date{\today}

\begin{abstract}
Large-area two-dimensional (2D) materials for technical applications can now be produced by chemical vapor deposition (CVD). Unfortunately, grain boundaries (GBs) are ubiquitously introduced as a result of the coalescence of grains with different crystallographic orientations. It is well known that the properties of materials largely depend on GB structures. Here, we carried out a systematic study on the GB structures in CVD-grown polycrystalline h-BN monolayer films by transmission electron microscope. Interestingly, most of these GBs are revealed to be formed via overlapping between neighboring grains, which are distinct from the covalently bonded GBs as commonly observed in other 2D materials. Further density functional theory (DFT) calculations show that the hydrogen plays an essential role in overlapping GB formation. This work provides an in-depth understanding of the microstructures and formation mechanisms of GBs in CVD-grown h-BN films, which should be informative in guiding the precisely controlled synthesis of large area single crystalline h-BN and other 2D materials.
\end{abstract}


\maketitle

\section*{INTRODUCTION}
Two-dimensional (2D) materials, such as graphene, hexagonal boron nitride (h-BN), and transition metal dichalcogenides (TMDs), etc. have gained great interest because of their remarkable and technologically useful properties\cite{novoselov_roadmap_2012,xu_graphene-like_2013,bhimanapati_recent_2015}. More interestingly, integrating these atomically thin crystals to heterostructures with a variety of properties opens up a new paradigm for nanoscale engineering\cite{geim_van_2013,novoselov_2d_2016}. Large-area, high-quality 2D materials are required for technical applications. Among numerous available synthesis methods, chemical vapor deposition (CVD) satisfies these demands and has been widely used to synthesis 2D materials because of merits in controllability and low cost\cite{li_large-area_2009,shi_synthesis_2010,lee_synthesis_2012}. However, large-area CVD-grown 2D materials are typically polycrystalline and therefore inevitably contain grain boundaries (GBs): the interfaces between differently oriented grains. Most studies on GBs in 2D materials were based on the covalently bonded GB (CBGB) model in which two grains are covalently connected by topological defects along the GBs\cite{Huang_Grains_2011,gibb_atomic_2013,cretu_evidence_2014,li_grain_2015,Zou_Predicting_2013}. Nevertheless, there is another type of GBs: overlapping GBs (OLGBs) formed by one grain climbing upon another, were also observed in 2D materials\cite{Tsen_Tailoring_2012,dong_formation_2017,Robertson_Atomic_2011,bayer_introducing_2017,van_der_zande_grains_2013,Najmaei_Vapour_2013} but gain much less attention. Properties of the polycrystalline 2D films are largely influenced by the constituent grains and the structure of GBs which largely depend on the misorientation angles between two grains\cite{Lee_High-Strength_2013,Wei_Nature_2012,Becton_Grain-Size_2015,mortazavi_modelling_2015,ding_effect_2014}. Hence, distribution of CBGBs and OLGBs in the film should also be considered because of the different impacts of OLGBs on the properties of materials as compared with CBGBs, i.e., mechanical strength or thermal conductivity. Though much work has been done so far to reveal the GB structures in 2D materials, there are very few studies revealing the distribution of different type of GBs in CVD-grown films as well as bridging the atomic scale and grain scales of GBs to get the full picture of GB structure. Actually, such knowledge is essential to fully understand the influence of GBs on the film properties of 2D materials.

A rich variety of GB configurations in monolayer h-BN can be identified due to the broken of inversion symmetry in the corresponding binary honeycomb lattice\cite{golberg_boron_2010,liu_dislocations_2012}, providing an excellent platform to systematically study the GBs in CVD-grown films. More importantly, h-BN itself exhibits extraordinary properties, such as high mechanical strength\cite{song_Large_2010} and good thermal conductivity\cite{ouyang_thermal_2010}, as well as excellent chemical stability\cite{golberg_boron_2010,chen_boron_2004}. It has been used in ultraviolet light emission\cite{kubota_deep_2007} and serves as a fundamental building block for van der Waals heterostructures because of its atomically smooth surface\cite{dean_boron_2010,xue_scanning_2011}. Thus, it’s essential to explore the GBs categories and distinguish the primary GB structures in CVD-grown h-BN. Furthermore, understanding the formation mechanism of GBs in h-BN could also help to understand the GBs in other 2D materials. 

Here, we build a full picture of GB structures in CVD-grown monolayer h-BN by conducting dark-field imaging analysis at the grain size scale, and spherical aberration-corrected high-resolution transmission electron microscopy (HRTEM) imaging to get the atomic structures. Our results show that OLGBs are found to be dominant in CVD-grown h-BN based on statistical analysis of over one hundred GBs. We also explored the structure of GBs formed between two rotationally aligned grains, which shows a folded structure feature instead of the perfect connection between two grains. To have a deep understanding on the formation mechanism of OLGBs in h-BN, we carried out density functional theory (DFT) calculations and revealed the energetic competition between CBGBs and OLGBs responsible for the formation of GBs, and the role of hydrogen during the formation of GBs.

\section*{METHODS}
\subsection{Synthesis of h-BN samples}
$Cu_{85}$$Ni_{15}$ alloy substrate was used for the CVD synthesis of h-BN domains as reported previously\cite{lu_synthesis_2015}. The introduction of Ni dramatically decreases the nucleation density of h-BN on the substrate and therefore increased the growing time of single crystals before they merge together, which results in large-size single crystalline grains. Ammonia borane ($H_{3}$$BNH_{3}$) was used as the precursor for the growth of h-BN samples. The growth was carried out in a two-zone-furnace. After the zone with $Cu_{85}$$Ni_{15}$ substrates reaches a temperature of 1070$^\circ \mathrm{C}$, the zone containing the ammonia borane was gradually heated to 75$^\circ \mathrm{C}$ in 10 min. During the growth, the chamber pressure was kept at 20-100 Pa under 100 sccm $H_{2}$. Normally, the growth of discrete h-BN domains takes about 60 mins while the growth of continuous h-BN film takes more than 90 mins. After the growth, the furnace was quickly cooled down to room temperature under the protection of Ar flow.

\subsection{TEM sample preparation and characterization}
As grown monolayer h-BN on a $Cu_{85}$$Ni_{15}$  substrate was spin coated with Poly(methyl methacrylate) (PMMA) and treated by oxygen plasma to remove the backside samples. And then the sample was transferred onto a TEM grid (quantifoil with a pore size of 1.2 $\mu$m) and ultrathin carbon film using electrochemical bubbling method\cite{wang_Electrochemical_2011}. Selected area electron diffraction (SAED) and HRTEM were done in a TEM (Titan G2 80-300, FEI) with a spherical aberration corrector on imaging side and a monochrometer which reduces the energy spread down to 0.12 eV. We used a positive spherical aberration ($C_{3}$) of 5 $\mu$m and a negative defocus of around -5.6 nm for atomic-resolution HRTEM. To reduce the radiation damage, this microscope was operated at an acceleration voltage of 80 kV and a  dose rate of about 1×$10^{6}$ e/$nm^{2}$s. All the HRTEM images were processed to remove the illumination variation by band-pass filter by Image J. Electron energy loss spectroscopy (EELS) experiments were conducted with Gatan Quantum 963 spectrometer in diffraction mode. SEM images were collected in a Hitachi SU70 at 3 kV.

\subsection{Density functional theory calculations}
First-principles density functional theory (DFT) calculations were carried out by using the Vienna ab initio simulation Package (VASP)\cite{kresse_ab_1993,kresse_efficiency_1996}. The interaction between the valence electrons and ion cores was treated by the projected augmented wave (PAW) method\cite{kresse_ultrasoft_1999}. The Perdew-Burke-Ernzerhof generalized gradient approximation (GGA) was chosen for the exchange-correlation interaction\cite{perdew_generalized_1996}. In addition, to describe the weak van der Waals interaction between h-BN layers and the substrate, the DFT-D2 method was used\cite{grimme_Semiempirical_2006}. The energy cutoff for the plane wave basis was set as 400 eV.

To model the structures of various h-BN edges and grain boundaries (GBs) on Cu substrate, a Cu(111) slab consisting of 3 atomic layers was constructed as the substrate, of which the unit cell size was 6.65×34.56×20.81 $Å^{3}$ and the bottom layer was fixed to mimic the bulk Cu foil. Due to the lattice mismatch between h-BN and Cu, the lattice constant of Cu substrate has already been reduced by less than 4\% to keep commensurate with h-BN. To model the formation of h-BN GBs, a pair of twin h-BN ribbons placed on Cu substrate was used. Their edges are either passivated by H or by the substrate. The k-point mesh was sampled as 5×1×1. All the structures were optimized until the force on each atom is within 0.01 eV/Å.

\begin{figure}
\includegraphics{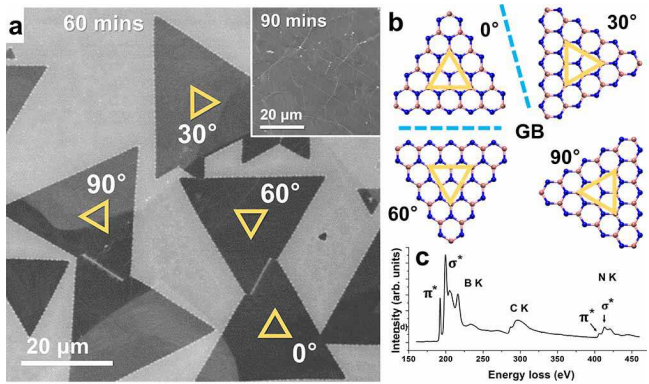}
\caption{\label{}(a) Secondary-electron microscopy (SEM) images of h-BN samples containing both discrete triangular grains and continuous film (inserted). The yellow triangles represent their crystallographic orientations. Note that 0\degree is an arbitrary value in relative to other orientations. (b) Schematic illustration of the orientations of the h-BN domains. The blue dash lines indicate the location of grain boundaries resulting from grains with two different orientations. Red balls represent B atoms and blue ones for N atoms. (c) The core-loss EEL spectra, where both B and N are identified. Note that the C K-shell signals are coming from the ultrathin carbon film served as the TEM supports.} 
\end{figure}

\begin{figure*}
\includegraphics{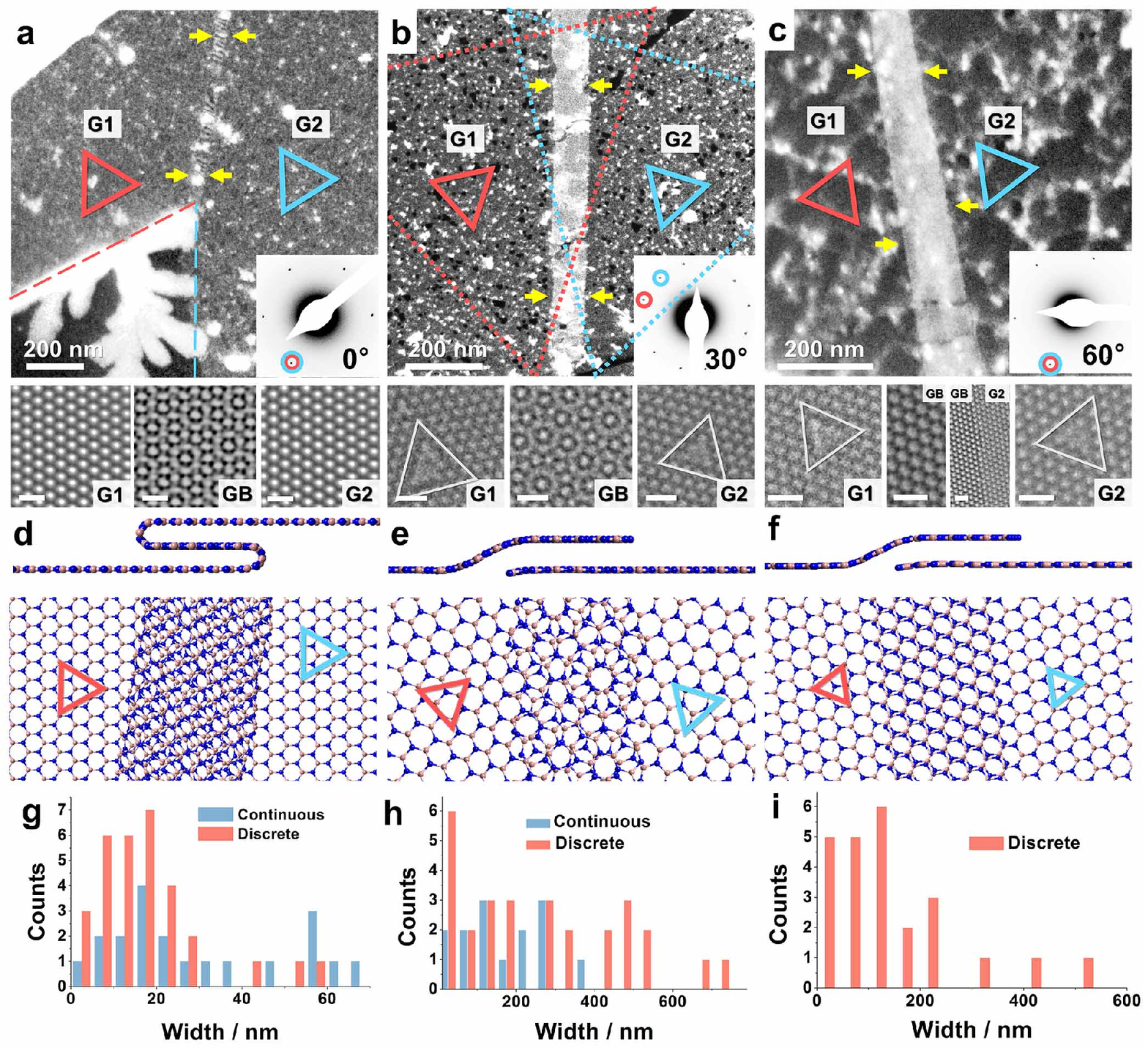}
\caption{\label{}(a-c) Dark-field TEM images of GBs with three typical misorientation angles: 0\degree, 30\degree (overlay of two DF-images) and 60\degree, with their SAED inserted. The triangles represent their orientations and colors indicate different grains named G1 (Grain 1) and G2 (Grain 2). The yellow arrows indicate the positions of GBs. Their corresponding HRTEM images are shown below. Note that the atomic images in Figure 2a are the inverse fast Fourier transform images. It should be noted that the diffraction spots of the interlayer in 0\degree-GB (insert in a) are invisible in SAED due to the rather weak small scattering intensity from the interlayer in the GB region. Scale bars for HRTEM images: 0.5 nm. (d-f) Cross-sectional (top) and planar (bottom) view of the corresponding structure model for GBs in (a-c). (g-i) Width distributions of three typical GBs. Due to the limited number of 60\degree-GBs observed in continuous films, the data of 60\degree-GBs in continuous films is not displayed here.} 
\end{figure*}

\section*{RESULTS AND DISCUSSION}

We adopt two types of h-BN samples grown on Cu–Ni alloy ($Cu_{85}$$Ni_{15}$) substrates with (100) dominated crystallographic facet at two different growth time (60 mins for discrete film and 90 mins for continuous film, respectively), as shown in Figure 1a. The triangular shaped h-BN grains displayed in Figure 1a indicate that these grains are single crystals with nitrogen-terminated zigzag edges\cite{liu_bn_2011}. Due to the 4-fold symmetry of the metal substrate and the epitaxial growth of h-BN on the substrate, there are four primary orientations of the h-BN grains\cite{wood_van_2015} as marked with yellow triangles in Figure 1a, leading to the formation of GBs with two orientations, as shown in Figure 1b. Besides the GBs with a misorientation angle of 30\degree, there is an inversion GB between two joint grains with a misorientation angle of 60\degree, owing to the polar structure of h-BN with alternating boron and nitrogen atoms in a honeycomb arrangement. Furthermore, it’s feasible to find the interface between two unidirectionally aligned h-BN grains in the samples, which is usually considered to be perfect, if the relative sliding of the two grains is ignored. Therefore, the samples used can serve as an ideal platform for a systematic study of the GBs in CVD-grown h-BN.

To fully reveal the GB structures in CVD-grown h-BN, information on both grain scale and atomic scale is necessary. The former one gives us the size, shape and edge orientations of grains as well as GB's morphology, while the latter one provides information of atomic structures. Herein, we use dark-field TEM (DF-TEM) and scanning electron microscope (SEM) to obtain the general picture of GBs, combined with atomically resolved HRTEM operating at 80 kV to get the detailed structures. The dark field images and corresponding atomic resolution images of three representative GBs are displayed in Figure 2a-c, with their atomic models shown in Figure 2d-f, respectively. The orientations of grains are determined by analyzing the selected area electron diffraction (SAED) patterns, combined with the triangular silhouette as illustrated in Figure 2a, and in together with the triangular holes with nitrogen terminated zigzag edges caused by electron beam with selected sputtering of boron atoms\cite{meyer_Selective_2009} as shown in atomic images of Figure 2b,c. Two independent sets of diffraction spots with a 30\degree rotation angle are observed in Figure 2b. In contrast, there is only one set of diffraction spots in 0\degree and 60\degree-GBs in Figure 2a,c, which indicates that two grains have the same or opposite orientations. Because the triangular silhouettes in Figure 2a show the same orientation, the corresponding GB would have a 0\degree  misorientation angle. In contrast, the two inverted triangular holes in shown Figure 2c indicate the inversion nature of two grains on two sides of the GB. 

\begin{figure}
\includegraphics{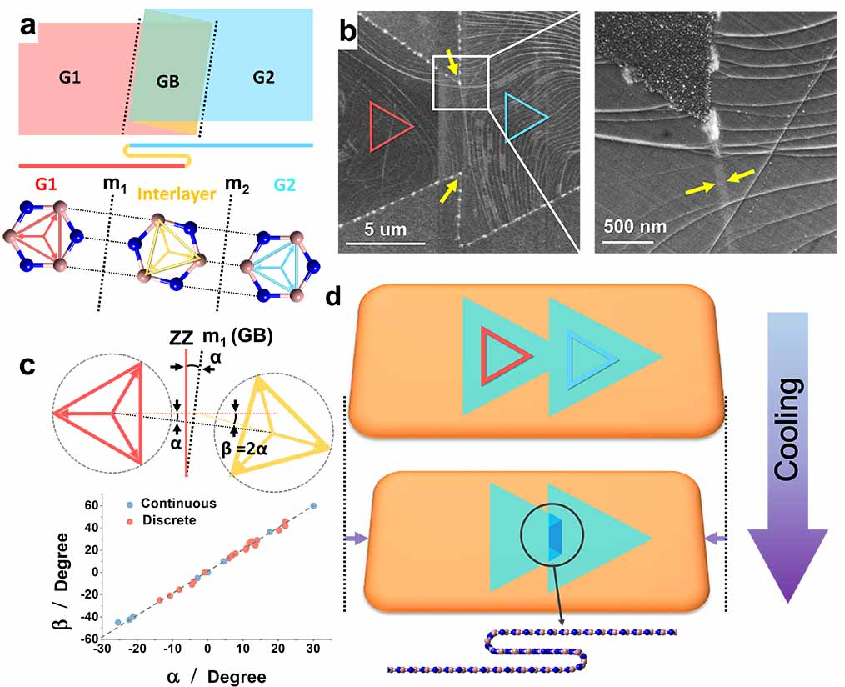}
\caption{\label{}(a) Schematic diagram of folded GBs in 0\degree and their structures of each layer with corresponding colors are shown at the bottom. The orange domain (the interlayer) is the reflection of red (blue) domain with respect to the folded line m1(m2) which is the GB direction. (b) SEM images of the 0\degree GBs. (c) Geometrical relationships of folded GBs with statistic plot of $\alpha$ and $\beta$ values in GBs of two samples shown below. The formula of the dashed line is  $\beta = 2\alpha$.  (d) The folding process of the 0\degree-GB GBs.}
\end{figure}

It is natural to propose that grains with the same orientation could coalesce and form a large single crystal with no line defects formed in between, if we ignore the relative sliding of the two grains\cite{li_growth_2016}. Here, we demonstrate that even for grains without any relative rotation, there still exist imperfections with a width of tens of nanometers between two grains and the corresponding atomic structure model is proposed as displayed in Figure 2d. The imperfections show a folded feature, which results in a relative rotation between the interlayer and other two layers. Furthermore, to verify the structure model in Figure 2d, we carry out a geometric analysis of the 0\degree-GB. If the structure of the 0\degree-GB is the same as sketched in Figure 2d and 3a, then the grains on two sides are the reflections of the interlayer with respect to the GB direction as indicated at the bottom of Figure 3a. It is not difficult to resolve the geometrical relationship from Figure 3c that $\beta = 2\alpha$, where $\alpha$ is the angle between the GB (marked by black dot line) and the zigzag boron direction (marked by red line) of the left grain, and $\beta$ is angle between the interlayer in the folded area and the left grain. Please refer to the APPENDIX B for detailed angle definition. By statistically analyzing the angle relationships of the data shown in Figure 3c, where $\beta$ is nearly two times of $\alpha$, we confirm the folded nature of 0\degree-GBs. 

As for the case of neighboring grains with misorientation angles of 30\degree and 60\degree, OLGB with a finite width is observed, as indicated by atomic-scale images in Figure 2b and 2c. Moiré patterns along the GB are clearly seen in Figure 2b due to the misorientation between two grains. The edges of the 30\degree-GB are relatively straight and nearly parallel with each other as revealed by the DF-TEM. However, they deviate from the direction of zigzag nitrogen terminated edges marked as triangles with dash lines in Figure 2b, which should be the growth front during CVD growth. As for neighboring grains with a misorientation angle of 60\degree, two edges with nitrogen termination should serve as the growth front of two grains, which are nearly atomic sharp and straight as seen in atomic scale HRTEM images in Figure 2c. The stacking order in the overlapping regions is assigned to be AA’ (B is sitting right above N), which is reported to be the most energetically favorable configureuration\cite{hod_graphite_2012}.

\begin{figure}
\includegraphics{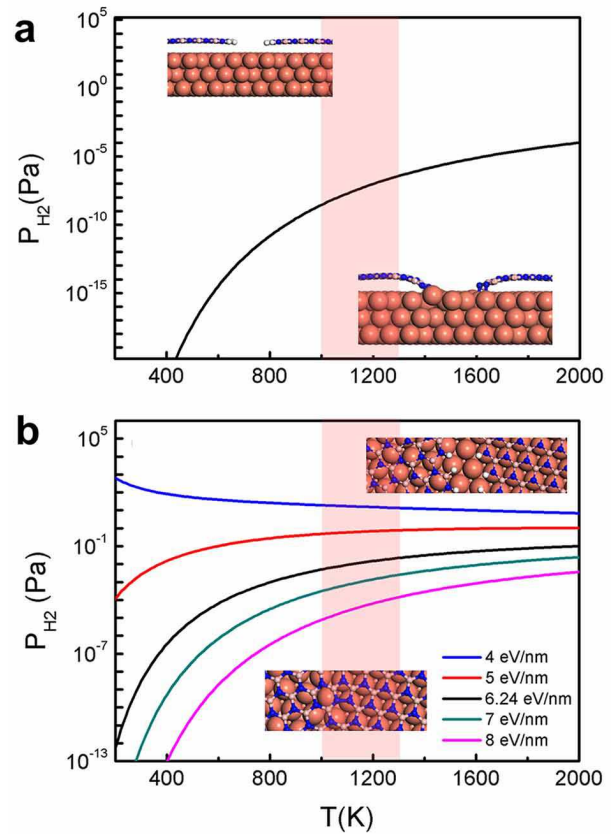}
\caption{\label{}(a) Thermodynamic diagrams between H-passivated h-BN edges and Cu substrate passivated h-BN edges. (b) Thermodynamic diagrams between H passivated h-BN edges and several covalently bonded GBs with different misorientation angles.} 
\end{figure}

In order to have a comprehensive understanding of the formation process of these different GBs, we counted the width of three types of GBs in both discrete and continuous h-BN films, as summarized in Figure 2g-i. Most of the 0\degree-GBs have a width of ~20 nm and the maximum width is less than 70 nm, while the width of the 30\degree and 60\degree-GBs  mostly ranges from 100 nm to 500 nm. The distinctions in structure and width distributions between 0\degree-GBs and 30\degree, 60\degree-GBs indicate that 0\degree-GBs have a different formation mechanism. It is inferred that the 0\degree-GB structure should be caused by the compressive stress concentration around the connecting region of two h-BN grains during the cooling process after the high-temperature growth stage of h-BN, as illustrated in Figure 3d. This is due to the mismatch of the thermal expansion coefficient between h-BN (-3.5×$10^{-6}$/K at 800 K)\cite{PhysRevB.63.224106} and the substrate (20×$10^{-6}$/K for Cu at 800 K)\cite{thomas_temperature_2015}, resulting in a strain in h-BN as large as $\sim$2$\%$ after cooling, supposing a growth temperature of 1000$^\circ \mathrm{C}$. Thus, the 0\degree-GBs are actually not GBs, but they may exist in most of the CVD-grown films, and certainly will influence the overall performance of those films.

There are no conventional CBGBs found in our data, in contrast to the cases of CVD-grown graphene and TMDs, where CBGBs account for a certain proportion\cite{dong_formation_2017,van_der_zande_grains_2013}. For the consideration of generality, we also checked the h-BN samples grown on different substrates, such as polycrystalline Cu and liquid Cu substrate, which are also widely used in CVD synthesis. The results we got don't show any obvious difference.

\begin{figure}
\includegraphics{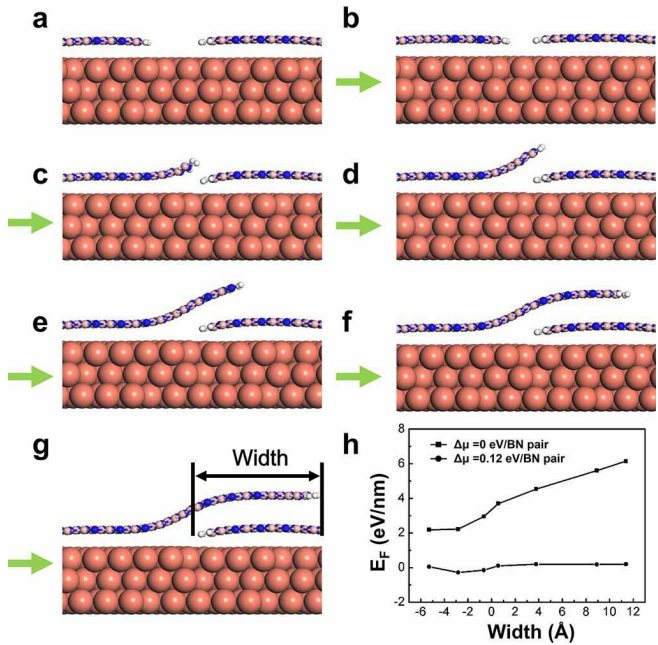}
\caption{\label{}(a-g) Formation process of overlapping GBs between two neighboring h-BN domains. (h) The formation energy evolution profile of (a-g).} 
\end{figure} 

\begin{figure}
\includegraphics{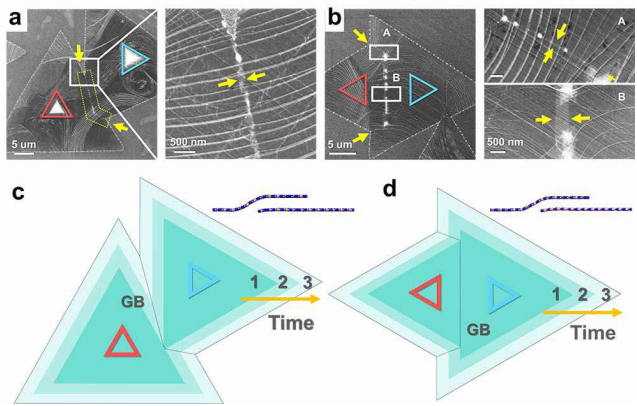}
\caption{\label{}(a,b) SEM images of 30\degree and 60\degree GBs. (c-d) Schematic diagram for h-BN growth mechanism. The graduated color grains indicate the grians at different grow time.}
\end{figure} 

To understand the formation mechanism of OLGBs, we further conduct DFT calculations. During the CVD growth process of h-BN, the growing edges should be either passivated by the metal atoms or by hydrogen atoms\cite{zhao_transition_2015,zhang_role_2014,wang_surface_2016,shu_edge_2014}. From our first-principles DFT calculations (Figure 4a), it is found that the required $H_{2}$ partial pressure for passivating the h-BN edge on the Cu substrate in the typical growth temperature range (1000-1300K) is quite low ($10^{-8}$-$10^{-6}$ Pa). Due to the low content of Ni in the employed Cu-Ni alloy substrate, the edges of h-BN on Cu-Ni alloy substrates during its CVD growth process would be mostly passivated by H atoms as in the case of Cu substrate, which is the pre-requisite for forming overlapping GBs. Furthermore, during the coalescence of neighboring h-BN grains, their edges would form either covalently bonded GBs by detaching H atoms or overlapping GBs through the climbing of one edge onto the other one, depending on their relative stabilities. Figure 4b shows the thermodynamic diagram between several covalently bonded h-BN GBs with different formation energies and H-passivated h-BN edges. Here, H-passivated h-BN edges were used to represent overlapping h-BN GBs. With the increase of the formation energy of h-BN covalently bonded GB from 4 to 8 eV/nm, the equilibrium $H_{2}$ partial pressure between overlapping h-BN GB and covalently bonded GB decreases sharply from 10 Pa to $10^{-4}$ Pa. Since both previous reports\cite{liu_dislocations_2012} and our calculations demonstrate that most of the calculated covalently bonded GBs exhibits a higher formation energy, (and/or near to) 4 eV/nm (the formation energy of GBs are discussed in APPENDIX C), we can conclude that, under a $H_{2}$ partial pressure of $\ge 10$ Pa, the majority of GBs in CVD h-BN would be overlapping GBs, consistent with our experimental findings. This may be the reason for such few experimental studies on CBGBs in monolayer h-BN at atomic scale\cite{gibb_atomic_2013,cretu_evidence_2014,li_grain_2015}. However, it should be noted that covalently bonded GBs can also be formed if their formation energies are small enough (small angle GBs) and the $H_{2}$ partial pressure is very low during the growth process.

Moreover, we studied the evolution of the formation energy during the process of an overlapping GB in h-BN. As shown in Figure 5, the formation process can be divided into three stages: (i) the approaching of edges from neighboring domains Figure 5a-b leads to little change of the formation energies, (ii) the climbing of one domain on another to form an overlapping structure (c-d) causes an increase of formation energy of about 2 eV/nm, (iii) the elongation of the overlapping region (e-g) results in a linear increase of the formation energy with a slope of about 2.3 $eV/nm^{2}$, due to the van der Waals interaction difference between h-BN layers and h-BN-Cu substrate. This increase of the formation energy can be easily compensated by a very small growth driving force (0.12 eV per B-N pair) of h-BN during the CVD process, suggesting that overlapping GBs in h-BN are easily formed.

Based on the mechanism explored above and the results of the grain-scale analysis, we can prove that the grain boundary formation is a kinetic process\cite{cheng_kinetic_2015}. As sketched in Figure 6c and 6d, two neighboring triangular grains firstly merged and continue to grow (as indicated by the graduated color), resulting in the formation of a boundary along intersections between two joint edges. The inclination of the GB then depends on the relative growth rates of the two grains. At micro-scale, after the encounter of two neighboring grains, the growth of the climbing grain will stop when the OLGB reaches a certain width, which eventually leads to the appearance of parallel edges in OLGBs and GB morphology as shown in the SEM images (Figure 6a,b). The strain induced by the cooling process will increase the width of the OLGBs, but won't be the cause of formation of OLGBs, due to the different magnitude of GB width between the 0\degree GBs and 30\degree (60\degree GBs) as illustrated in Figure 2g-i. The width of 30\degree-GBs doesn't increase with growth time, as the average width in continuous films is not larger than that in discrete films (Figure 2h). We infer two possible mechanisms of such width limited process. The first mechanism is that the overlapping layer grows between the catalyst (substrate) and the growing h-BN grain, and the feeding of precursor comes through the edges of the upper h-BN layer\cite{kidambi_situ_2014}, where the width of the OLGBs corresponding to the diffusion length of the precursor. The second mechanism is that after the overlapping layer grows onto another layer, the growth stops due to the absence of catalysis by the substrate. Further experiments and theoretical analysis are necessary to really understand such phenomenon and its mechanism.

\section*{CONCLUSION AND OUTLOOK}
Our results demonstrate that the dominant GB structure in CVD-grown h-BN is overlapping GBs (and folded GBs for aligned grains), rather than covalently bonded GBs. This gives an entirely different view of GBs in h-BN and 2D material family, and provides a model system for researchers to further explore the effects of GBs on properties. Since CVD is one of the most suitable methods for massive production of h-BN, such OLGBs should be paid more attention and be considered when 2D membranes are used in industrial applications. 

From previous discussions, we could propose that we can change the GB structure by tuning the $H_{2}$ pressure. Moreover, when h-BN forms in-plane heterostructures with other 2D materials such as graphene, if they nucleate independently from each other, the edges of h-BN/graphene will be passivated either by H or by metal substrates, and thus results in different interfaces. This can explain the formation of overlapping heterostructure between h-BN and graphene as previously observed\cite{kim_synthesis_2013}. However, the actual role of $H_{2}$ in CVD process could be more complex than we expected. For instance, $H_{2}$ has a great impact on the growth speed and morphology of h-BN domains, and even has etching effect to h-BN edges\cite{sutter_chemical_2011,wang_growth_2015,stehle_anisotropic_2017}. Thus, tuning the GB structures by either tuning $H_{2}$ or other methods should be systematically  studied further. In addition, the energetic competition between CBGBs and OLGBs should exist in all 2D materials, including transition metal dichalcogenides (TMDs). Though there is no hydrogen during some synthesis of TMDs, OLGBs are likely to form when OLGBs are much energy favorable compared with CBGBs. 

We prove that wrinkling should be considered during the synthesis of large-area single-crystalline h-BN films by coalescence of numerous grains grown from aligned h-BN nuclei. In fact, the mismatch of thermal expansion coefficient between materials and substrates are ubiquitous in all the 2D materials, which would easily lead to the folding process in the region between two grains or in inside the grain. The structure of the connection region in other materials should be similar to the structure we presented here. Jang et al.demonstrated wafer-scale wrinkle-free multilayer h-BN growth on a sapphire substrate by LPCVD\cite{jang_wafer-scale_2016}, due to the small difference in the thermal expansion coefficients between h-BN and sapphire. Deng et al. proved that graphene is wrinkle-free grown on single crystal Cu(111) thin film due to that Cu(111) is the lowest energy surface with the smallest thermal expansion among other crystallographic orientations, and larger interaction of graphene with Cu(111) than that of Cu(100) and Cu(110)\cite{deng_wrinkle-free_2017}.

In summary, we have shown that overlapping GBs are the primary structure in CVD grown h-BN. And there is folded structure in two rotational aligned grains. The theoretical study shows that the edges of h-BN on CVD growth are dominantly passivated by H atoms, which is a precondition of forming overlapping GBs. And an energetic competition between covalently bonded GB and overlapping GBs are systematic theoretical studied. This study reveals the GB nature in CVD-grown h-BN and gives the mechanism behind the phenomenon, which will allow us to precisely control the growth of desired h-BN and other 2D materials.

\section*{ACKNOWLEDGMENTS}
\begin{acknowledgments}
 The authors thank Prof. Bin Wu, Zhepeng Zhang and Prof. Yanfeng Zhang for providing us high-quality h-BN samples grown on liquid Cu and polycrystal Cu substrates for cross-check, and thank Xiaowei Wang, Xujing Ji and Fenfa Yao for critical comments on editing the manuscript. This work was financially supported by the National Science Foundation of China under Grants 51772265, 51472215, 51761165024 and 61721005, the National Basic Research Program of China under Grants 2014CB932500 and 2015CB921004, and the 111 project under Grant B16042. The work on electron microscopy was carried out at the Center of Electron Microscopy of Zhejiang University. The work done in Korea was supported by Institute for Basic Science (IBS-R019-D1). The work performed at the Shanghai Institute of Microsystem and Information Technology, Chinese Academy of Sciences, was partially supported by the National Key R\&D program (Grant No. 2017YFF0206106), the National Science Foundation of China (Grant No. 51772317, 11604356), the China Postdoctoral Science Foundation (Grant No. 2017M621563, 2018T110415), the Science and Technology Commission of Shanghai Municipality (Grant No. 16ZR1442700, 18511110700) and Shanghai Rising-Star Program (A type) (Grant No.18QA1404800). The work done in Nanjing was supported by National Natural Science Foundation of China (51535005, 51472117), the Research Fund of State Key Laboratory of Mechanics and Control of Mechanical Structures (MCMS-I-0418K01, MCMS-I-0418Y01, MCMS-0417G02, MCMS-0417G03), the Fundamental Research Funds for the Central Universities (NP2017101, NC2018001), and a Project Funded by the Priority Academic Program Development of Jiangsu Higher Education Institutions.
\end{acknowledgments}
X.R. and J.D. contributed equally to this work.


\section{APPENDIX A: The definition of misorientation angle}
The misorientation angle $\theta$, is defined as the angle between two zigzag boron directions of left (red dash line) and right (blue dash line) grains, ranging from -60\degree to 60\degree. $\theta$ is positive if the vector from red line (left grain) to blue line (right grain) shows a clockwise rotation as in shown in Figure 7.
It should be noted that 30\degree-GB in the main text refer to both -30\degree-GB and +30\degree-GB following the definition mentioned above, while the 60\degree-GB only refers to 60\degree as defined here (-60\degree-GB is barely found in our experiments).

\begin{figure}
\includegraphics{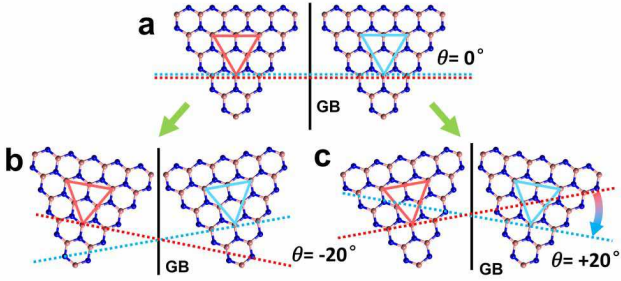}
\caption{\label{} The definition of misorientation angle $\theta$.} 
\end{figure}

\section{APPENDIX B: Detailed structure determination of grain boundaries}

\begin{figure}
\includegraphics{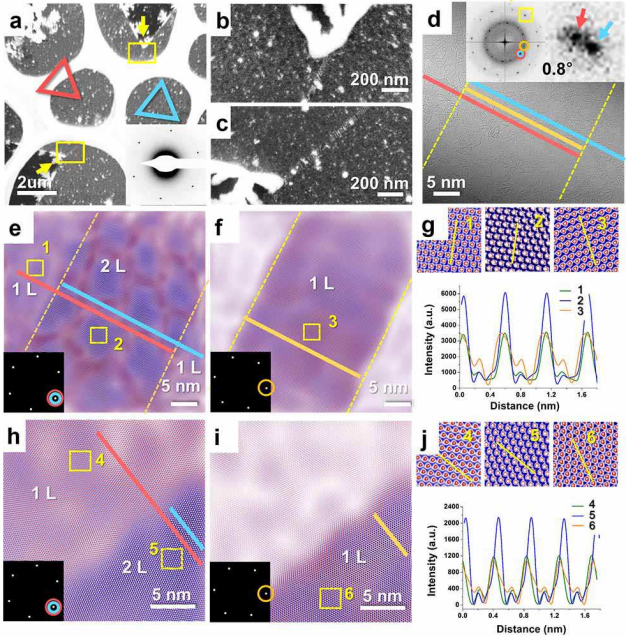}
\caption{\label{}(a) Dark-field image of 0\degree-GBs, which is the same as Figure 2a in main test (the image in the main text is rotated 40\degree in anticlockwise direction for better display). (b,c) enlarged images in a which are marked by orange rectangles. (d) HRTEM image of Figure b and its corresponding FFT. The spot in red and blue circles is from the left and right domains, while the spot marked by the orange circle are from the interlayer. There are small misorientation angle(0.8\degree) between left and right domain as shown in the high order FFT spots (e,f) inverse FFT of the information of domains and the interlayer. Inserted: cut mask in FFT. (h,i) inverse FFT of HRTEM images in c. Inserted: cut mask in FFT (g,j) the intensity profile of the corresponding region marked with numbers, which indicate the layer number of the certain region.} 
\end{figure}

\begin{figure}
\includegraphics{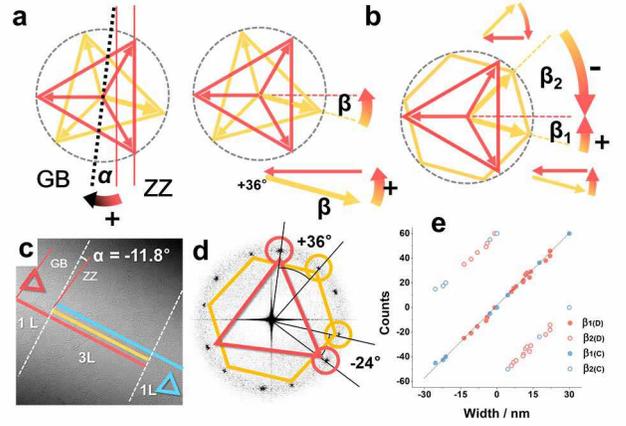}
\caption{\label{}(a) The definition of $\alpha$ and $\beta$ (b) Schematic diagram of the orientations of the interlayer (we can only get the orientation of hexagonal lattice in the experiments) and the left grain, which shows two $\beta$ values. (c) HRTEM image of 0\degree GB which is the same as Figure 8d. Its corresponding FFT is displayed in d. (e) The relationships between $\alpha$ and $\beta$ ($\beta_{1}$ and $\beta_{2}$ ). D represent discrete sample and C represent continuous sample} 
\end{figure}

\begin{figure}
\includegraphics{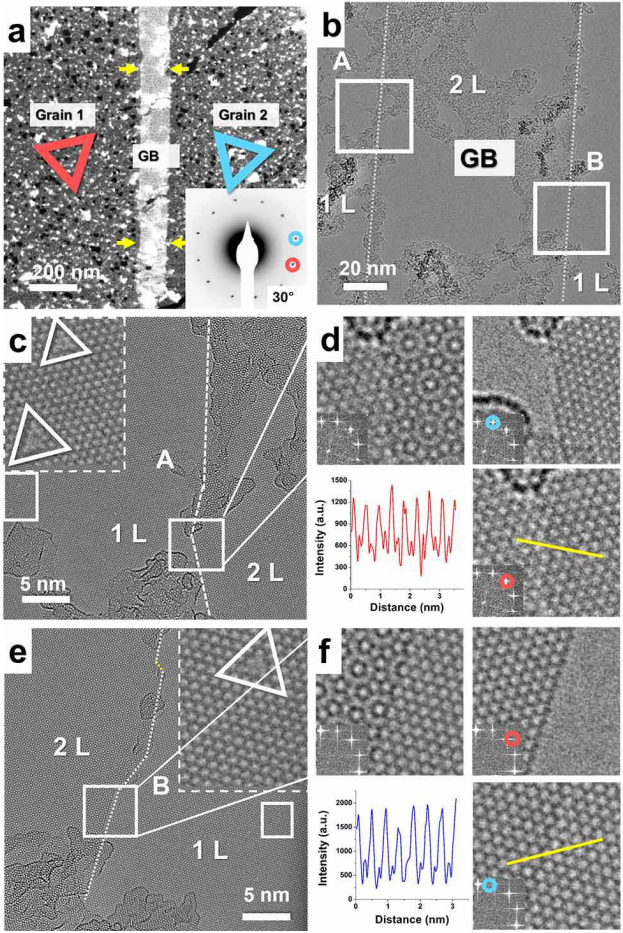}
\caption{\label{}Structure of 30\degree GB (a) Dark-field image of 30\degree GB (the same as Figure 2b in main text). (b) HRTEM image of low magnification, and detail structures of local areas marked as A and B are displayed in c and e. (d,f) Inverse FFT of local region in c and e, with their masks inserted and intensity profile in certain region, which indicate the layer numbers.} 
\end{figure}

\begin{figure}
\includegraphics{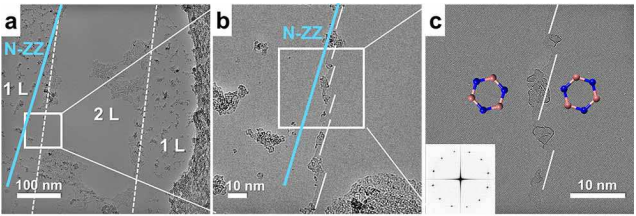}
\caption{\label{}(a) HRTEM image of 30\degree-GBs. The blue line indicate the nitrogen terminated zigzag edge. (b,c) Local structures in a, which shows a step-like edge with straight edges terminated by nitrogen.} 
\end{figure}

\begin{figure}
\includegraphics{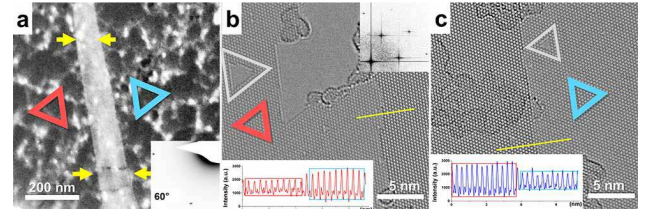}
\caption{\label{}(a) Dark field image of 60\degree-GBs (the same as Figure 2c in main text). (b,c) the left and the right region of grain boundary and the internsiy profile indicate the layer number of the certain region. } 
\end{figure}

The data in Figure 2a and Figure 8 were acquired in a discrete sample on the ultrathin carbon film (to keep the triangle shape of h-BN grains). We can not directly see the atomic structure from the images. However, there is a strong signal in FFT as seen in Figure 8d, which make it easy to determine the orientations in the FFT and the layer number in inverse FFT images. Note that in inverse FFT images, it’s improper to explore the detailed structures such as defect structures due to the artifacts resulting from the filtering process. The moiré patterns in Figure 8b, 8c,e are due to a small misentation angle between the left and right grains in local area.

The 0\degree GBs are formed due to the folding process during the cooling, and are actually not GBs but wrinkles. Such folding process could also be happened inside the grain. Thus we can not determine whether the line defects in continuous film (lost the triangular shape during grain coalescence) are the GBs between two grains or the wrinkles inside the grains. Nevertheless, the structures in continuous film also follow the geometrical relationships as shown in Figure 3b, which indicate that they should be the same as the structure of 0\degree GBs in discrete sample.

Detailed angle definition in geometric analysis is displayed in Figure 9. $\alpha$ is the angle between the GB line (marked by the black dot line) and the zigzag boron direction (marked by red line) of the left grain, which range from -30\degree to 30\degree owing to the symmetry of h-BN. And $\alpha$ is positive if the black line (GB) is in the clockwise direction from the red line. $\beta$ is an angle of the interlayer (yellow) with respect to the left grain and accord with the vector relation, ranging from -60\degree to 60\degree. $\beta$ is positive if the vector shows an anticlockwise rotation. 

In the experiments, we can get the orientation of the left and right grains, and the orientation of the hexagonal lattice of the interlayer based on the FFT. However, we can not distinguish the boron or nitride atoms in the hexagonal lattice of the interlayer. Therefore, the orientation of the interlayer is represented as hexagon which results in two $\beta$ values, $\beta_{1}$ and $\beta_{2}$, as shown in Figure 9b and Figure 9d (+36\degree and -24\degree). The value of -24\degree is around two times of the angle of GB with respect to zigzag boron direction of the left grain (-11.8\degree) as shown in Figure 9c. The statistical analysis of $\alpha$ and two $\beta$ values($\beta_{1}$, $\beta_{2}$) is shown in Figure 9e, which indicates that only one $\beta$ approximately satisfy the formula ($\beta$=2$\alpha$).

The edges of the 30\degree-GB are relatively straight and nearly parallel with each other viewed from the DF-TEM image (Figure 2b) and low-magnification TEM image (Figure 11a), whose direction deviate from the direction of nitrogen terminated zigzag edges marked by blue line (Growth front during CVD growth). However, when we look carefully at the edge, we found that the edge are formed by nitrogen terminated steps, which further indicate the kinetic nature of GB formation.

\section{APPENDIX C: Computational details}
\subsection{Calculation of the formation energies of different types of h-BN edge on Cu substrate}
The formation energy of H-passivated h-BN edge is first calculated, which is defined by the following equation:
\begin{equation} 
E_{F-H}=((E_T-E_S-N\times\varepsilon_{BN}-N_H\times\varepsilon_{H2}/2)/4l_{H2}/2)/4l
\end{equation}
where $E_T$ and $E_S$ are the energies of the total system and the substrate, respectively. N and $N_H$ are the number of BN pairs and H atoms, respectively. $\varepsilon_{BN}$ and $\varepsilon_{H2}$ are the energies of a BN pair in h-BN on Cu substrate and $H_2 molecule$, respectively. L is the width of the unit cell.

The formation energy of substrate passivated h-BN edge is thus calculated by:
\begin{equation} 
E_{F-S}=((E_T-E_S-N\times\varepsilon_{BN}-N_H\times\varepsilon_{H2}/2)/l-2×E_{F-H})/2
\end{equation}
and the formation energy of a covalently bonded GB in h-BN on Cu substrate is
\begin{equation} 
E_{F-GB}=(E_T-E_{S}-N\times\varepsilon_{BN}-N_H\times\varepsilon_{H2}/2)/l-2\times E_{F-H}
\end{equation}

\subsection{Calculation of the formation energy evolution during the overlapping of two h-BN domains}
The formation energy during the overlapping of two neighboring h-BN domains is calculated by:
\begin{equation}
\begin{aligned}
 E_F=&(E_T-E_{S}-N\times\varepsilon_{BN}-N_H\times\varepsilon_{H2}/2)/l-2E_{F-H}\\
&-N\times\Delta\mu_{BN}
\end{aligned}
\end{equation}
where $\Delta\mu_{BN}$ is the chemical potential difference between a BN pair in h-BN on Cu substrate and in growth precursor.

\subsection{Calculation of the chemical potential of H2 and thermodynamic diagrams}
The hydrogen chemical potential $\mu_{H}$ in $H_2$ gas at a temperature T and a pressure P is calculated by\cite{landau_statistical_1980}:
\begin{equation} 
\mu_H=[E_{H2}-k_BTln(gk_BT/P\times\xi_{trans}\xi_{rot}\xi_{vib})]/2
\end{equation}
where $E_{H2}$ is the energy of a $H_{2}$ molecule, $k_{B}$ is the Boltzmann constant, g equals 2 accounting for the degree of degeneracy of the electron energy level, $\xi_{trans}$, $\xi{rot}$ and $\xi{vib}$ are the partition functions of translation, rotation and vibration motions.
The thermodynamic diagrams between H passivated h-BN edges and covalently bonded grain boundary or Cu passivated h-BN edges are obtained by calculating their Gibbs free energy difference $\Delta G$ as:
\begin{equation} 
\Delta G=\Delta E_{T}+\Delta F_V-N_{H}\times \mu_{H}
\end{equation}
where $\Delta E_{T}$ is the total energy difference between H passivated h-BN edges and covalently bonded grain boundary or Cu passivated h-BN edges. $\Delta F_{V}$ is the vibration entropy difference between the H passivated h-BN edges and covalently bonded grain boundary or Cu passivated h-BN edges.

\subsection{Calculation of the formation energy of covalently bonded h-BN GBs as a function of its mis-orientation angle.}
To calculate the formation energy of covalently bonded GBs in h-BN as a function of its mis-orientation angle, $\theta$, twin GBs in h-BN are chosen.
The GB formation energy is defined as:

\begin{equation} 
E_{F-GB0}=(E_{T}-N\times\varepsilon_{BN0}-N_{H}\times\varepsilon_{H2}/2)/l-2\times E_{F-H}
\end{equation}

where $\varepsilon_{BN0}$ is the energy of BN pair in suspended h-BN. The calculated GB formation energy profile is shown in Figure S 17.

\begin{figure}
\includegraphics{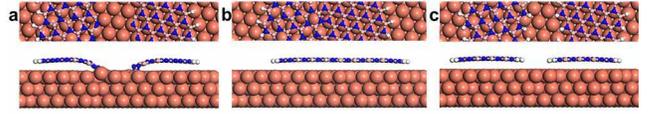}
\caption{\label{}(a) h-BN nanoribbons on Cu substrate with both H-passivated edges and substrate passivated edges, (b) h-BN nanoribbon on Cu substrate with H passivated edges and a twin GB, (c) h-BN nanoribbons on Cu substrates with only H passivated edges.} 
\end{figure}

\begin{figure}
\includegraphics{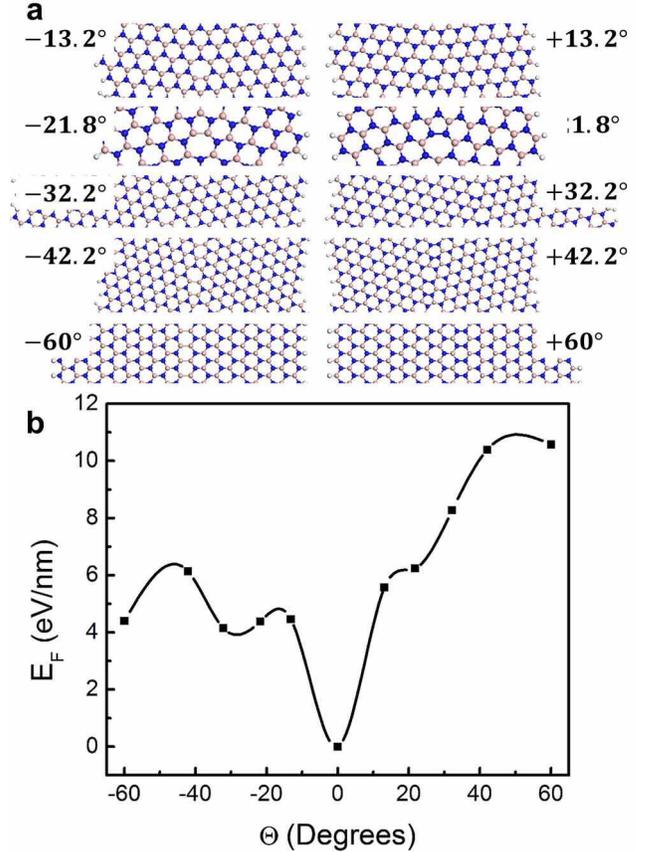}
\caption{\label{}Structures of covalently bonded GBs in h-BN with different mis-orientation angles and their corresponding formation energy} 
\end{figure}


\bibliography{apssamp.bib}

\end{document}